\documentclass[preprint,fleqn,aps,pra,nofootinbib,groupedaddress]{revtex4-2}
\usepackage{graphicx}
\usepackage{amsmath}
\usepackage{amssymb}
\usepackage{hyperref}

\begin{document}

\title{Molecular effects in low-energy muon transfer from muonic hydrogen to oxygen}

\author{I. Boradjiev$^{1,2}$}
\author{A. Adamczak$^3$}
\author{D. Bakalov$^1$}
\author{M. Baruzzo$^4$}
\author{R. Benocci$^{5,6}$}
\author{M. Bonesini$^{5,7}$}
\author{S. Capra$^{8,9}$}
\author{E. Christova$^1$}
\author{M. Clemenza$^{5,7}$}
\author{M. Danailov$^{4,10}$}
\author{P. Danev$^1$}
\email{Corresponding author: petar\_danev@abv.bg}
\author{E. Fasci$^{11,12}$}
\author{L. Gianfrani$^{11,12}$}
\author{A.D. Hillier$^{13}$}
\author{K. Ishida$^{14}$}
\author{A. Menegolli$^{15,16}$}
\author{E. Mocchiutti$^4$}
\author{S. Monzani$^{4,17}$}
\author{L. Moretti$^{11,12}$}
\author{C. Pizzolotto$^4$}
\author{R. Rossini$^{15,16}$}
\author{A. Sbrizzi$^{18}$}
\author{M. Stoilov$^1$}
\author{H.S. Tonchev$^1$}
\author{A. Vacchi$^{4,17}$}
\author{E. Vallazza$^5$}
\author{K. Yokoyama$^{13}$}

\affiliation{$^1$ Institute for Nuclear Research and Nuclear Energy, Bulgarian Academy of Sciences, boul. Tsarigradsko ch. 72, Sofia 1784, Bulgaria} 
\affiliation{$^2$ University of Architecture, Civil Engineering and Geodesy, Boul. H. Smirnenski 1, Sofia 1046, Bulgaria}
\affiliation{$^3$ Institute of Nuclear Physics, Polish Academy of Sciences, Radzikowskiego 152, PL31342, Krak\`{o}w, Poland}
\affiliation{$^4$ INFN Sezione di Trieste, via A. Valerio 2, Trieste, Italy}
\affiliation{$^5$ INFN Sezione di Milano Bicocca, Piazza della Scienza 3, Milano, Italy}
\affiliation{$^6$ Dipartimento di Scienze dell'Ambiente e della Terra,   Universit\`{a} di Milano Bicocca, Piazza della Scienza 1, Milano,  Italy}
\affiliation{$^7$ Dipartimento di Fisica G. Occhialini, Universit\`{a} di Milano Bicocca, Piazza della Scienza 3, Milano, Italy}
\affiliation{$^8$ INFN Sezione di Milano, Via Celoria 16, Milano, Italy}
\affiliation{$^9$ Dipartimento di Fisica, Universit\`{a} degli Studi di Milano, Via Celoria 16, 20133 Milano}
\affiliation{$^{10}$ Sincrotrone Elettra Trieste, SS14, km 163.5, Basovizza, Italy}
\affiliation{$^{11}$ INFN Sezione di Napoli, Via Cintia, Napoli, Italy}
\affiliation{$^{12}$ Dipartimento di Matematica e Fisica, Universit\`{a} della Campania ``Luigi Vanvitelli'', Viale Lincoln 5, Caserta, Italy}
\affiliation{$^{13}$ ISIS Neutron and Muon Source, STFC Rutherford-Appleton Laboratory, Didcot, OX11 0QX, United Kingdom}
\affiliation{$^{14}$ Riken Nishina Center, RIKEN, 2-1 Hirosawa, Wako, Saitama 351-0198, Japan}
\affiliation{$^{15}$ INFN di Pavia, Via A. Bassi 6, Pavia,  Italy}
\affiliation{$^{16}$ Dipartimento di Fisica, Universit\`{a} di Pavia, via A. Bassi 6, Pavia, Italy}
\affiliation{$^{17}$ Dipartimento di Scienze Matematiche, Informatiche e Fisiche, 
Universit\`{a} di Udine, via delle Scienze 206, Udine, Italy}
\affiliation{$^{18}$ INFN Sezione di Bologna, viale Berti Pichat 6/2, Bologna, Italy}

 \date{\today}
 
 \begin{abstract}
In the present study we determine 
from the available experimental data the cross section of muon transfer to molecular oxygen at low energies with account of the oxygen molecule structure. 
Building on an earlier work, the results highlight the role of the molecular structure effects and significantly improve the agreement with theoretical calculations of the muon transfer rate. 
An efficient computational model of the kinetics of processes involving muonic hydrogen atoms in gaseous mixture of H$_2$ and O$_2$ is developed and analyzed. The model is applied in the description of the FAMU experiment for the measurement of the hyperfine splitting in muonic hydrogen and the Zemach radius of the proton.
 \end{abstract}

\maketitle

\section{Introduction}

The reaction of muon transfer from muonic hydrogen to oxygen
  \begin{equation}
  p\mu^-+O_2\rightarrow p+(O\mu^-)+O
  \label{eq:transfer}
  \end{equation}
has been attracting the attention of both experimentalists and theorists since the discovery of the double-exponential time spectra of muonic oxygen X-rays \cite{mulhauser}.
The interest in the subject was revived recently in  relation to the projects to measure the hyperfine splitting in the ground state of muonic hydrogen \cite{jinst18,epja,japs-las,
crema-las,crema-new} by laser spectroscopy,   and determine out of it the electromagnetic   Zemach radius of the proton \cite{ourZemach,cjp}.
The experimental method of the FAMU collaboration  \cite{epja}, in particular, 
consists in stopping negative muons in a  mixture of H$_2$ and O$_2$, exciting the muonic hydrogen atoms 
$p\mu$ to the triplet spin state with a pulsed tunable laser, and determining the resonance frequency for the singlet-to-triplet hyperfine transition from the response of the time distribution of the muon transfer events (\ref{eq:transfer}) in the target to variations of the laser wavelength.
Modeling the time distribution and optimization of the FAMU experiment requires the detailed knowledge of the collision energy dependence of the rate 
of the reaction (\ref{eq:transfer}) in the thermal and near epithermal range. 
Several advanced calculations of this rate have been reported  \cite{dupays1,cdlin,tcherbul,romanov22},
which needed, however, experimental verification. In a recent work \cite{ours} the muon transfer rate to oxygen was determined for energies below 0.1 eV 
from experimental data obtained in thermalized mixed H$_2$ and O$_2$ target at fixed temperatures in the range between 70 K and 336 K  \cite{pla20,pla21}.
The derivation in \cite{ours} completely neglected the effects of the structure of the molecule O$_2$ and was based on the assumption that the energy distribution of the muonic hydrogen atoms is Maxwellian. In the present paper we substantially upgrade these results.
In Sect.~\ref{sec:II} we formulate a model of the FAMU experiment in terms of a Boltzmann-Lorentz type kinetic equation for the energy and spin distribution of thermalized $p\mu$ atoms propagating in hydrogen-oxygen gas mixture. The time distribution of $\mu$-transfer events is expressed in terms of the $p\mu$ energy distribution and the unknown rate of the process of muon transfer.   
In Sect.~\ref{sec:IIIa} we extract the muon transfer cross section and rate in collisions of $p\mu$ atoms with O$_2$ molecules from FAMU experimental data with account of the internal degrees of freedom of O$_2$.
The computational techniques applied in \cite{ours} are upgraded to reduce the model-dependence of the results and make them applicable in processing the experimental data on muon transfer to other nuclei, e.g. carbon \cite{monzani}. 

\section{Modeling the FAMU experiment}
\label{sec:II}

\subsection{A brief overview of the FAMU experimental method}

The FAMU experimental method exploits the anticipated (since the early observations in Ref.~\cite{werth0}) and later rigorously established (in \cite{pla20,pla21,ours}) strong energy dependence of the rate of muon transfer from hydrogen to oxygen. 
The muonic hydrogen atoms, while propagating in the gaseous mixture of hydrogen and oxygen, collide with oxygen that leads to the reaction (\ref{eq:transfer}); the events of muon transfer are signaled by the characteristic X-rays emitted during the fast de-excitation of the muonic oxygen initially formed in highly excited states. 
A laser pulse of near-resonance frequency, irradiating the gaseous target, will excite (fraction of) the $p\mu$ atoms to the triplet spin state; after de-excitation in subsequent collisions with the surrounding 
H$_2$ molecules these atoms acquire additional kinetic energy of nearly 0.12 eV. As long as the rate of muon transfer varies with the $p\mu$ kinetic energy, the observed time distribution of the characteristic X-rays is perturbed as compared to the time distribution in absence of laser radiation. The resonance laser frequency is  recognized by the  response of the X-ray time distribution to variations of the laser wave length. (For details see Refs.~\cite{nimb12,jinst16,jinst18,epja}).
The time distribution of the characteristic X-rays
is the main observable in the FAMU experimental method, and the goal of the present work is to contribute to the development of a reliable model for its evaluation, needed for the optimization and data analysis of the FAMU experiment.

\subsection{Modeling the time distribution of the muon transfer events}
\label{sec:2B}

The muonic hydrogen atoms $p\mu$ are formed in excited states when negative muons are stopped in the hydrogen target. At the FAMU experimental conditions (a few bar pressure and temperatures not exceeding 336 K) they are promptly de-excited to the ground state.
We restrict our consideration to time intervals (at least 150 ns after the muon stop \cite{jinst18}) for which, on the ground of Monte Carlo simulations, we can safely assume that all hydrogen muonic atoms are in the $1S$ Coulomb state, in either the singlet ($F=0$) or triplet ($F=1$) spin state.
Our quantitative model of the propagation of muonic hydrogen atoms in gaseous mixture of hydrogen and oxygen is a Boltzmann-Lorentz type kinetic equation \cite{boltz3} for the energy and spin state distribution of $p\mu$ atoms in the ground 1S state, with a few simplifying assumptions, valid at FAMU experimental conditions: (1) the probability for either elastic or inelastic scattering of $p\mu$ by O$_2$ is negligible compared to the anomalously high probability for muon transfer (\ref{eq:transfer}); (2) collisions between the extremely rare $p\mu$ atoms are neglected; (3) the probability of formation of $pp\mu$ is negligible. 

Denote by $n_F(E;T;t)\,dE$ the probability that at time $t\ge0$ a hydrogen muonic atom in the 1S-state, propagating in thermalized H$_2$ and O$_2$ mixture at temperature $T$, is in the total spin state $F$ with lab frame kinetic energy in the interval 
$(E,E+dE)$. 
Collisions with H$_2$ molecules may change the energy $E$ and the spin state $F$ while,  with a good accuracy, scattering by O$_2$ may be neglected compared to the much faster process of muon transfer (\ref{eq:transfer}). Other reactions to take into account are the muon transfer to deuterium  and the muon decay. Thus, $n_F(E;T;t)$ satisfies the equation
\begin{align}
&\frac{dn_F(E;T;t)}{dt}=-\lambda^{\rm(dis)}(E;T)\,n_F(E;T;t)
\label{eq:integdif}\\
&+S(t)\,(n_{(1-F)}\left(E;T;t)-n_F(E;T;t)\right)
\nonumber
\\
&+\phi\,c_H\sum_{F'}\int\limits_0^{\infty}n_{F'}(E';T;t)\,{\cal M}_{F'F}(E',E;T)\,dE'
\nonumber\\
&-\phi\,c_H\sum_{F'} n_{F}(E;T;t)\int\limits_0^{\infty}{\cal M}_{FF'}(E,E';T)\,dE'.
\nonumber
\end{align}
Here 
$\lambda^{\rm(dis)}(E;T)$ denotes the overall disappearance rate of the $p\mu$ atoms
\begin{align*}
&\hspace*{-0.8cm}\lambda^{\rm(dis)}(E;T)=\lambda^0+\phi\,c_d\lambda^{(pd)}(E;T)
+\phi\,c_O\lambda^{(pO)}(E;T),
\end{align*}
$\phi$ is the number density of the gas mixture in liquid hydrogen density (LHD) units $\rho_{\rm LHD}=4.25\times10^{22}$ cm$^{-3}$, $c_H$, $c_O$, and $c_d$ are the atomic concentrations of the hydrogen and oxygen components and the deuterium impurities satisfying 
$c_H+c_O+c_d=1$, $\lambda^0=0.455162\,10^6$ s$^{-1}$ is the free muon decay rate, and 
$\lambda^{(pO)}(E;T)$ and $\lambda^{(pd)}(E;T)$ are the rates of muon transfer to  oxygen (in reaction (\ref{eq:transfer})) and deuterium (in collision with a D nucleus) at target temperature $T$ from a $p\mu$ atom with lab frame kinetic energy $E$, normalized to LHD \cite{jinst18}. According to \cite{jacot-pd}, in the energy range of interest the rate of muon transfer to deuterium can be assumed to be independent of the energy: 
$\lambda^{(pd)}(E;T)=\overline{\lambda^{(pd)}}=1.64\,10^{10}$ s$^{-1}$. 
$S(t)$ in the second term is the probability per unit time for spin-flip 
$F\rightleftarrows (1-F)$ of $p\mu$ stimulated by the laser radiation. 
Finally,
${\cal M}_{FF'}(E,E')\,dE'$ is the probability per unit time that 
an incoming $p\mu$ atom in spin state $F$ and with kinetic energy $E$ is scattered by H$_2$ 
into an outgoing state with spin $F'$ and kinetic energy in the interval $(E',E'+dE')$. 
The number $dn_X(T;t)$ of characteristic X-ray photons emitted in the interval $(t,t+dt)$,  which is equal to the number of muon transfer events to oxygen in the same interval, is given by:
\begin{align}
dn_X(T;t)=dt\sum_F\int dE\,\phi c_O\lambda^{(pO)}(E;T)\,n_F(E;T;t),
\label{eq:timespect}
\end{align}
therefore the time spectrum of the characteristic X-rays is determined by the time evolution of the kinetic energy distribution $n_F(E;T;t)$ of $p\mu$ and the energy and temperature dependence $\lambda^{(pO)}(E;T)$ of the rate of muon transfer to oxygen. The rest of this subsection is focused on the evaluation of $n_F(E;T;t)$; the muon transfer rate will be considered in Sect.~\ref{sec:2.1}.

An efficient approach to resolve Eq.~(\ref{eq:integdif}) is the multigroup method \cite{cacuci}: the range of kinetic energies is truncated to a properly selected cut-off value $E_{\rm max}$ and split to a set of $n$ intervals 
$(\epsilon_{i-1},\epsilon_i),i=1,...,n$, with $\epsilon_0=0, \epsilon_n=E_{\rm max}$. Define
\begin{align}
&E_i=(\epsilon_{i-1}+\epsilon_i)/2,\ \   
\lambda^{(pO)}_i(T)=\lambda^{(pO)}(E_i;T),
\label{eq:forSM}\\
&n_{Fi}(T;t)=\int\limits_{\epsilon_{i-1}}^{\epsilon_i} n_F(E;T;t)\,dE,
\nonumber\\ 
&M_{\rm Fi,F'i'}(T)=\int\limits_{\epsilon_{i'-1}}^{\epsilon_{i'}}\!
{\cal M}_{FF'}(E_i,E';T)\,dE'. 
\nonumber
\end{align}
Eq.~(\ref{eq:integdif}) is then transformed into a system of linear differential equations 
\begin{align}
&\frac{dn_{Fi}(T;t)}{dt}=
S(t)\big(n_{(1-F)i}(T;t)-n_{Fi}(T;t)\big)
\label{eq:discr-det}\\
&-(\lambda^0+\phi\,c_d\,\overline{\lambda^{(pd)}}+\phi\,c_O\,\lambda^{(pO)}_i(T))n_{Fi}(T;t)
\nonumber\\
&+\phi\,c_H\sum_{F'i'}\big(n_{F'i'}(T;t)\,M_{F'i',Fi}(T)
\nonumber\\
&-n_{Fi}(T;t)M_{Fi,F'i'}(T)\big).
\nonumber
\end{align}
which is equivalent to (\ref{eq:integdif}) in the limit $E_{\rm max}\to\infty,\ n\to\infty$.
In matrix form Eq.~(\ref{eq:discr-det}) reads 
$\mathbf{\dot{n}}(T;t)=\mathbf{A}(T)\,\mathbf{n}(T;t)$, where
\begin{align}
&A_{Fi,F'i'}(T;t)=S(t)\,\delta_{(1-F)i,F'i'}-S(t)\,\delta_{Fi,F'i'}
\label{eq:matrform}\\
&-(\lambda^0+\phi\,c_d\,\overline{\lambda^{(pd)}}
\!-\!\phi\,c_O\,\lambda^{(pO)}_i(T))\,\delta_{Fi,F'i'}
\nonumber\\
&+\phi\,c_H\big(
M_{F'i',Fi}(T)\!-\!\delta_{Fi,F'i'}\sum_{F''i''}M_{Fi,F''i''}(T)\big)
\nonumber
\end{align} 
with formal solution
\begin{align}
\mathbf{n}(T;t)=\exp \left(t\,\mathbf{A}(T;t)\right)\,\mathbf{n}(T;0),
\label{eq:closeform}
\end{align}
where $\mathbf{n}(T;0)$ is the initial energy and spin distribution of the muonic hydrogen atoms.
Two sets of elastic scattering rate matrices $M_{Fi,F'j}(T)$, calculated in \cite{adamczak-codes} and 
corresponding to truncation energy $E_{\rm max}=100$ eV and binning into $n=201$ and $n=385$ intervals, have been used in our analysis. Within the required precision the results are independent of the truncation energy and the selected energy bins. To facilitate the use of Eq.~(\ref{eq:closeform}) in modeling tasks we include in the Supplemental material \cite{suppl} the matrices $M_{Fi,F'j}(T)$ with 385 energy bins for $T=$80K. 

\begin{figure}
\begin{center}
\includegraphics[scale=0.6]{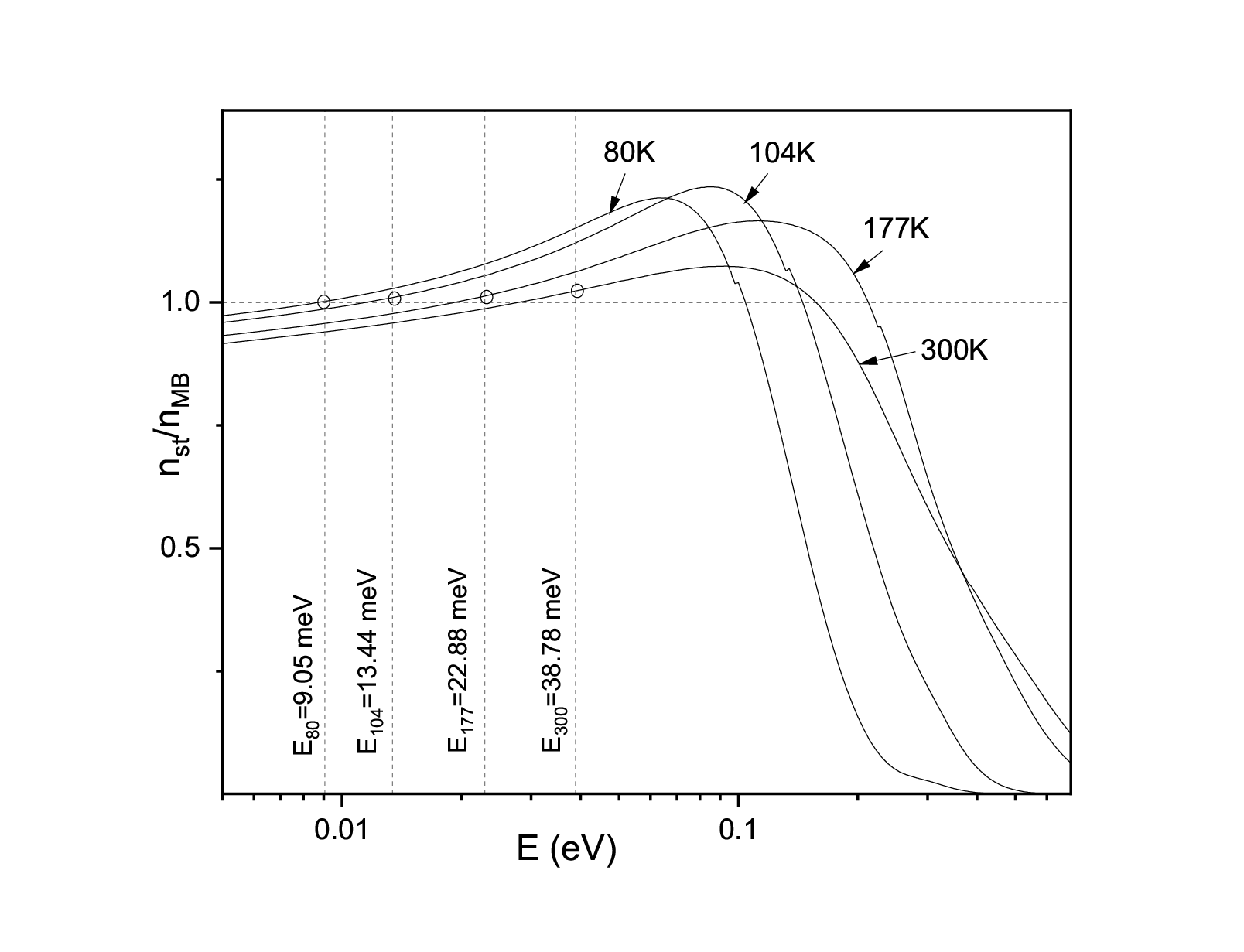}
\caption{The ratio of the stationary energy distribution of the muonic hydrogen atoms to the 
Maxwell-Boltzmann energy distribution $\mathbf{n}^{\rm(st)}(T)/\mathbf{n}^{\rm MB}(T)$ for 
H$_2$-O$_2$ gas target temperatures $T=$80, 104, 177, and 300 K. The values of the ratios at the mean thermal energies $E_T=3k_BT/2$, marked with circles, turn out to be very close to 1. Near epithermal states are overpopulated at the expense of subthermal and higher energy states.}
\label{fig:ratios}
\end{center}
\end{figure}

The closed form solution of Eq.~(\ref{eq:closeform}), while less useful in the general case of time-dependent matrix $\mathbf{A}(T;t)$, turns out to be an efficient tool in modeling the physical processes investigated in the FAMU experiment in absence of laser radiation, i.e. when $S(t)=0$. Among other, it allows to evaluate the stationary energy distribution 
$\mathbf{n}_{\rm st}(T)$ of the $p\mu$ atoms, established after collisions with the surrounding H$_2$ molecules have erased the memory of the initial energy distribution at the time of formation of $p\mu$; it is enough to follow the time evolution of $\mathbf{n}(T;t)$ until it converges to a steady state.\footnote{If the lifetime of $p\mu$ were infinite one might evaluate 
$\mathbf{n}_{\rm st}(T)$ as the eigenvector of $\mathbf{A}(T)$ corresponding to the zero eigenvalue of 
$\mathbf{A}(T)$. Both approaches lead to numerically identical results.} 
The time needed to reach this stationary energy distribution depends on the gaseous target temperature and pressure. In various FAMU experimental conditions it varies approximately between 100 and 1000 ns.
Fig.~\ref{fig:ratios} shows the ratio of $\mathbf{n}_{\rm st}(T)$ to the 
Maxwell-Boltzmann distribution 
$\mathbf{n}_{\rm MB}(T)=\{2\sqrt{E_i/\pi(k_BT)^3}\exp(-E_i/k_BT), i=1,...,n\}$ 
for a few temperatures in the range $70K\le T\le 300K$.  The distribution 
$\mathbf{n}_{\rm st}(T)$ deviates from the Maxwell-Boltzmann distribution: states with higher energy are depopulated in favor of the near-epithermal states. 
The stability of the results obtained with the two alternative binning schemes described above and by varying the truncation energy $E_{\rm max}$, as well as the convergence to the Maxwell-Boltzmann distribution in test runs using the scattering cross sections of $p\mu$ by protons \cite{melezh}, show that the deviation of 
$\mathbf{n}_{\rm st}(T)$ from $\mathbf{n}_{\rm MB}(T)$
is not a numerical artifact. 
It may not be related to the oxygen admixture either, because the simulations in pure hydrogen and 
in the H$_2$--O$_2$ mixture of used in FAMU measurements ($\phi<0.05, c_O\sim2\,10^{-4}$, 
\cite{pla20,pla21}) give very close results.
We conclude that the deviation is due to the inelastic collisions of $p\mu$ with H$_2$, in which the atom transfers part of its kinetic energy to rotational excitation of the molecule. The phenomenon may be of interest in the study of a variety of low-energy processes involving hydrogen muonic atoms. The Supplemental material \cite{suppl} gives the numerical values of the ratio of the two distributions and a smooth fit to them. In the evaluation of the molecular effects on the muon transfer rate to oxygen in Sect.~\ref{sec:2.1}, however, the deviation between the two distributions is shown to have negligible effects. 

\subsection{Equation for the cross sections of muon transfer to oxygen with account of the molecular structure}
\label{sec:2.1}

Knowledge of the energy and temperature dependence of the muon transfer rate to oxygen,
$\lambda^{\rm (pO)}(E;T)$, is fundamental to complete the evaluation of the time distribution of the characteristic X-rays in  Eq.~(\ref{eq:timespect}). 
In what follows we derive an equation that relates the muon transfer cross section to the experimentally observable rate of process (\ref{eq:transfer}) with account of the effects of the structure of the oxygen molecule. The numerical solution of the equation and the analysis of the results in the framework of the FAMU project are discussed in Sect.~\ref{sec:III}.

The definitions and the notations of the investigated functional dependence need   clarification. 
The rate of the elementary three-body process $p\mu+{\rm O}^{8+}\to {\rm O}^{8+}\mu+p$ is denoted by $\lambda(E)$ and depends on a single argument -- the collision energy $E$ in their center-of-mass (CM) reference frame or, equivalently, on the relative velocity of the colliding species; the same notation is used when this does not lead to ambiguity.
The rate $\lambda^{(pO)}(E;T)$ in Eq.~(\ref{eq:timespect}) refers to muon transfer process in which the oxygen is in thermal equilibrium at temperature $T$, while the $p\mu$ lab frame kinetic energy is an independent variable. 

In the considerations in Ref.~\cite{ours} 
the observable rate of muon transfer at fixed target temperature $T$, $\Lambda(T)$, was expressed in terms of the muon transfer rate  
$\lambda$ as function of the relative velocity 
$\mathbf{v}_{AM}\!=\!\mathbf{v}_A\!-\!\mathbf{v}_M$ of $p\mu$ and the O$_2$ molecule:
\begin{align*}
&\hspace*{-.7cm}\Lambda(T)=
\int d^3\mathbf{v}_A \,f_A(\mathbf{v}_A;T)\int d^3\mathbf{v}_M \,
f_M(\mathbf{v}_M;T)\,\lambda(v_{AM}),
\end{align*}
where 
$\mathbf{v}_A$ and 
$\mathbf{v}_M$ denote the lab frame velocities of the interacting $p\mu$ atom and the O$_2$ molecule, respectively, $v_{AM}=|\mathbf{v}_M-\mathbf{v}_A|$,
$f_A(\mathbf{v}_A;T)$ and $f_M(\mathbf{v}_M;T)$ are the corresponding velocity distribution densities at temperature $T$ (assumed in \cite{ours} to be Maxwellian).  
The above expression for $\Lambda(T)$ corresponds to the simplifying assumption that the oxygen nuclei are ``frozen'' at the center-of-mass of O$_2$, which does not provide the accuracy needed for the reliable modeling of the FAMU experiment. 
To account for the internal degrees of freedom of the oxygen molecule, we now distinguish the translational thermal motion of O$_2$ from the vibrational and rotational motion of the oxygen nucleus involved in the three-body muon transfer process by taking 
$\Lambda(T)$ in the form
\begin{align*}
&\Lambda(T)\!=\!\rho_M\!\! 
\int \!d^3\mathbf{v}_A\!\int\! d^3\mathbf{v}_M\!\int\! d^3\mathbf{v}_N\,
v_{AM}\,\sigma(|\mathbf{v}_N\!-\!\mathbf{v}_A|)
\\
&\times
f_A(\mathbf{v}_A;T)\,f_M(\mathbf{v}_M;T)\,f_N(\mathbf{v}_N\!-\!\mathbf{v}_M;T),
\end{align*}
where $\rho_M$ is the number density of oxygen atoms in the gas target, 
$\mathbf{v}_N$ is the lab frame velocity of the oxygen nucleus, 
$f_N(\mathbf{v}_N\!-\!\mathbf{v}_M;T)$ is its velocity distribution in the O$_2$ center-of-mass rest frame, and 
$\sigma(|\mathbf{v}_N-\mathbf{v}_A|)$ is the cross section of muon transfer from $p\mu$ to O, expressed as function of their relative velocity. The rate 
$\lambda(v)$ and the cross section $\sigma(v)$ are related by $\lambda(v)=\rho_M\,v\,\sigma(v)$.
The change of variables 
\begin{align*}
&\mathbf{x}=\mathbf{v}_A,\ 
\mathbf{v}=\mathbf{v}_N-\mathbf{x},\ 
\mathbf{u}=\mathbf{v}_M-\mathbf{x},\\
&\mathbf{v}_A=\mathbf{x},\ 
\mathbf{v}_M=\mathbf{u}+\mathbf{x},\ 
\mathbf{v}_N=\mathbf{v}+\mathbf{x}
\end{align*}
transforms the above expression into
\begin{align}
&\Lambda(T)=\rho_M \int d^3\mathbf{x}\int d^3\mathbf{v}\int d^3\mathbf{u}\,
|\mathbf{u}|\,
\sigma(v)
\label{eq:LAMN3D1}\\
&\times f_A(\mathbf{x};T)\,f_M(\mathbf{x}+\mathbf{u};T)\,f_N(\mathbf{v}-\mathbf{u};T).
\nonumber
\end{align}
By introducing the notation
\begin{align*}
&g(\mathbf{v};T)=\!
\int d^3\mathbf{u}\,|\mathbf{u}|\,
\left(\int d^3\mathbf{x} \, f_A(\mathbf{x};T)\,f_M(\mathbf{x}\!+\!
\mathbf{u};T)\right)
\\
&\times f_N(\mathbf{v}-\mathbf{u};T)
\end{align*}
Eq.~(\ref{eq:LAMN3D1}) takes the form
\begin{align}
&\Lambda(T)=\rho_M \int d^3\mathbf{v}\,
\sigma(v)\,
g(\mathbf{v};T).
\label{eq:basic3}
\end{align}
which is the 3-dimensional (3D) integral equation for the unknown cross section 
$\sigma(v)$, analogous to Eq.~(5)  of Ref.~\cite{ours}. 
In fact, all of the above velocity distribution densities are isotropic: they depend only on the magnitude, but not on the direction of their vector argument. We label with an overline the corresponding 1D distribution densities of the vector magnitude: 
\begin{align*}
\overline{f_A}(x;T)=4\pi x^2\,f_A(|\mathbf{x}|;T),
\end{align*}
normalized with 
$\int_0^{\infty}dx\,\overline{f_A}(x;T)=1$,
and similar for $\overline{f_{M}}$, $\overline{f_{N}}$, and $\overline{g}$. 
The 1D distribution $\overline{g}$ corresponding to $g$ is
\begin{align}
&\overline{g}(v;T)=\frac{v}{2}
\int\limits_0^{\infty}\!du\,\overline{f_{AM}}(u;T)
\int\limits_{|v-u|}^{v+u}ds\,\frac{\overline{f_{N}}(s;T)}{s},
\label{eq:ofAMN}
\end{align}
where
\begin{align}
&\overline{f_{AM}}(u;T)=\frac{u}{2}
\int\limits_0^{\infty}\!dx\,\frac{\overline{f_{A}}(x;T)}{x}
\!\int\limits_{|u-x|}^{u+x}\!dt\,\frac{\overline{f_{M}}(t;T)}{t}
\label{eq:ofAM}
\end{align}
and the 1D integral equation for the cross section $\sigma(v)$ takes the form 
\begin{align}
&\Lambda(T)=\rho_M\int_0^{\infty}dv\,\sigma(v)\,\overline{g}(v;T).
\label{eq:basic1}
\end{align}

\section{Cross section of the muon transfer from muonic hydrogen to oxygen}
\label{sec:III}

\subsection{Numerical evaluation of the cross section}

\label{sec:IIIa}

The kernel $\overline{g}(v;T)$ is expressed in terms of the distributions densities 
$\overline{f_{A}}$, $\overline{f_{M}}$, and $\overline{f_{N}}$. The velocity distribution of the oxygen molecules is Maxwellian:
\begin{align}
&\overline{f_{M}}(t;T)=\sqrt{\frac{2}{\pi}}
\left(\frac{m_M}{k_BT}\right)^{3/2} t^2\exp\left(-\frac{m_Mt^2}{2k_BT}\right).
\end{align}
The velocity distribution of the muonic hydrogen atoms $\overline{f_{A}}(x;T)$ 
can be expressed in terms of an interpolation $\widetilde{n}_{\rm st}(E;T)$ of 
the discrete stationary distribution density $\mathbf{n}_{\rm st}$ of Sect.~\ref{sec:2B}: 
\begin{align}
&\overline{f_{A}}(x;T)=m_A\,x\,\widetilde{n}_{\rm st}(E;T)\vert_{E=m_A x^2/2},
\label{eq:nst}
\end{align}
where the interpolating function $\widetilde{n}_{\rm st}(E;T)$ satisfies
\begin{align}
&\widetilde{n}_{\rm st}(E_i;T)=\mathbf{n}_{{\rm st},i}, i=1,...,n.
\end{align}
The velocity distribution of the interacting nucleus of O$_2$ in the oxygen molecule CM 
frame, $\overline{f_{N}}(s;T)$, was evaluated using the results on the momentum distribution density of the nuclei of a diatomic molecule with account of the temperature-dependent population of the excited ro-vibrational states from Ref.~\cite{pettitt}; the range of physical parameters (temperature, density, etc.) of relevance for the FAMU experiment were tested and shown to satisfy the conditions for the validity of the approximations in Ref.~\cite{pettitt}. The contribution of the ground vibrational and lowest 40 rotational excitations of O$_2$ has been  accounted for in 
$\overline{f_{N}}(s;T)$. 
The change in the population of the rotational states of O$_2$ due to the interaction with the scattered muonic atom were shown to be negligible (see Appendix 1).  
The convolution integrals in Eqs.~(\ref{eq:ofAMN},
\ref{eq:ofAM}) were evaluated numerically for velocities up to $v_{\rm max}=$12000 m/s. 
It is worth noting that in spite of the noticeable deviation of $\overline{f_{A}}(x;T)$ from the Maxwell-Boltzmann distribution of the $p\mu$ velocity (see Fig.~\ref{fig:ratios}), 
$\overline{f_{AM}}(u;T)$ is very close to the Maxwell-Boltzmann distribution of the relative velocity $u$ of $p\mu$ and O$_2$. The explanation is that the main contribution to the integral over $x$ in (\ref{eq:ofAM}) comes from the range of thermal energies where the above deviation is small.

\begin{figure}
\begin{center}
\includegraphics[scale=0.6]{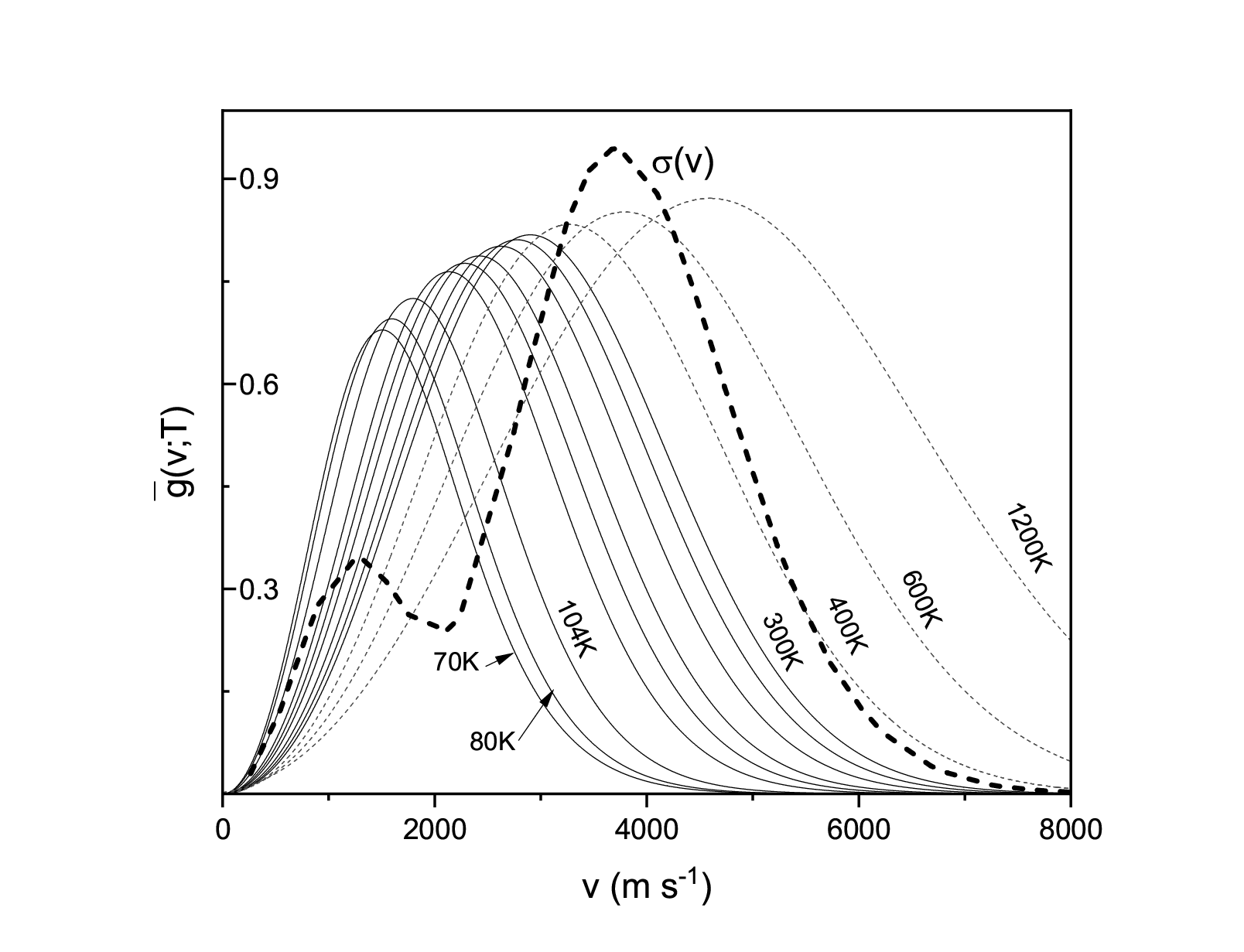}
\caption{The kernels $\overline{g}(v;T)$ for the experimentally investigated temperature range, calculated with the $p\mu$ velocity distribution $\overline{f_A}$ of Eq.~(\ref{eq:nst}) (solid lines), and extrapolated for higher temperatures (short-dashed lines). 
The thick dashed line represents the profile of the muon transfer cross section 
$\sigma(v)$.}
\label{fig:gauss}
\end{center}
\end{figure}

\begin{figure}[b]
\begin{center}
\includegraphics[scale=0.6]{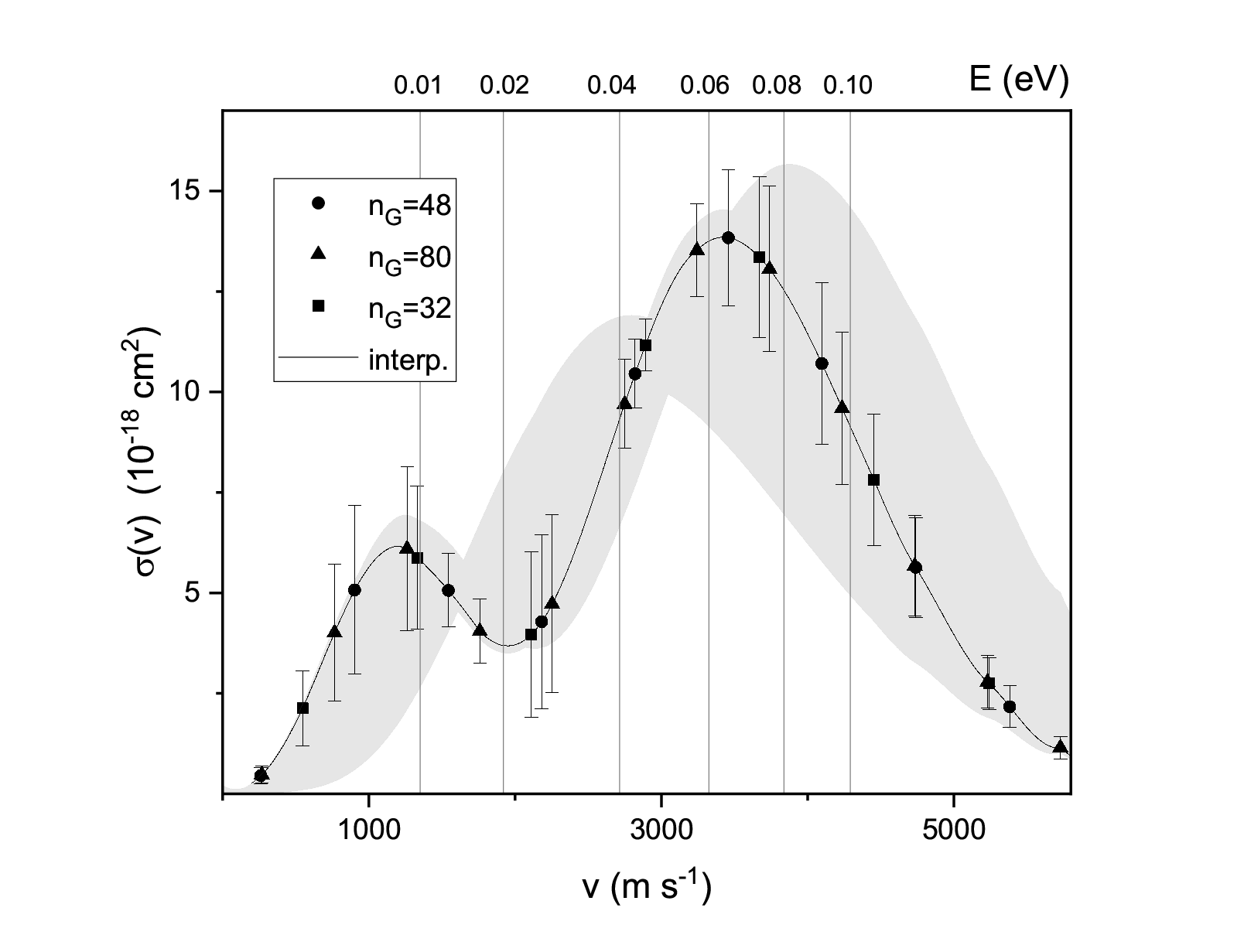}
\caption{Values of the cross section $\sigma(v)$ of muon transfer from a $p\mu$ atom to an oxygen nucleus, calculated from Eq.~(\ref{eq:linsys}) for a discrete set of relative velocities, corresponding to the nodes of Gauss-Hermite quadratures of rank $n_G=32$, 48, and 80. 
The error bars represent the statistical error due to the experimental uncertainty of the FAMU data. The solid line is a cubic interpolation of the discrete data points.
The shadowed area illustrates the uncertainty related to the lack of data at higher temperatures. The vertical lines link the lower (velocity) and upper (energy) scales.}
\label{fig:symbols}
\end{center}
\end{figure}

To obtain the numerical values of the muon transfer cross section $\sigma(v)$, we resolve Eq.~(\ref{eq:basic1}) by applying the technique already successfully used in \cite{ours}.
The integral in (\ref{eq:basic1}) is discretized by using an appropriate Gauss quadrature of rank $n_G$ with nodes $x_n$ and weights $w_n, n=1,...,n_G$. (On the go we substitute 
$\rho_{\rm LHD}$ to $\rho_M$ since the data on $\Lambda(T)$ are normalized to LHD):
\begin{align}
&\Lambda_k=\Lambda(T_k)=\rho_{\rm LHD}\sum\limits_{n=1}^{n_G}\sigma(x_n)\,w_n\,g(x_n;T_k)
\end{align} 
and solve the the resulting regularized linear system for the discrete values of 
$\sigma_n=\sigma(x_n)$:
\begin{align}
&\sum\limits_{n=1}^{n_G} C_{k,n}\sigma_n=\Lambda_k,k=1,...,10,
\label{eq:linsys}
\\
&\quad C_{k,n}=\rho_{\rm LHD}\,w_n\,g(x_n;T_k).
\nonumber
\end{align}
For the temperatures explored by the FAMU collaboration, the kernel 
$\overline{g}(v;T)$ has Gaussian-like shape (see Fig.~\ref{fig:gauss}), and the Hermite-type quadratures \cite{abram} are the best choice. To guarantee that the numerical error in evaluating the integrals in (\ref{eq:basic1}) with a single quadrature does not exceed $10^{-5}$  for any $T$ in the range 70K$<T<$336K, we use quadratures of sufficiently high rank $n_G=$ 32, 48, and 80, centered at $v_c=2500$ m/s and FWHM of $s_c=4000$ m/s: 
$x_n=v_c+s_c\,x_n^{\rm(T)}$, $v_n=s_c\,w_n^{\rm(T)}$, where $x_n^{\rm(T)}$ and $w_n^{\rm(T)}$ are the standard Hermite quadrature nodes and weights corresponding to weight function $\exp(-x^2)$. 
Eq.~(\ref{eq:linsys}) is regularized using the truncated singular value decomposition method 
\cite{kaipio} by decomposing the matrix $C$ in the form $C=UDV^T$ with orthogonal $U,V$ and diagonal $D$, and keeping (similar to \cite{ours}) the first $n_T=3$ diagonal matrix elements of $D$. Denote by $\sigma_i^{\rm(n_G)},i=1,...,n_G$, the solutions of the regularized linear system (\ref{eq:linsys}). The solutions 
$\sigma_i^{\rm(n_G)}$ for which $x_i^{\rm(n_G)}<0$ or $x_i^{\rm(n_G)}>v_{\rm max}$ prove to be several orders of magnitude smaller than the solutions with abscissas within the interval 
$(0,v_{\rm max})$, and are discarded. The collection of valid solutions are shown on 
Fig.~\ref{fig:symbols}. The subsets calculated with $n_G=$ 32, 48, and 80 are well aligned with each other; we assume that the smooth interpolation of the discrete points on the plot  
represents our estimate of the relative velocity dependence of the cross sections $\sigma(v)$ of muon transfer in collisions of $p\mu$ with an oxygen nucleus. 

The independence of $\sigma(v)$ on the quadrature rank and the insensitivity to variations (in reasonable limits) of the quadrature parameters $v_c$ and $s_c$ confirm the stability of the applied numerical techniques. The error bars give a conservative estimate of the statistical uncertainty of the calculated cross sections due to the experimental uncertainty of the input data $\Lambda_k,k=1,...,10$. 
 
We are not aware of any method to rigorously evaluate the uncertainty of $\sigma(v)$ due to the finiteness of the investigated temperature range, and can only qualitatively estimate it. 
Most important appears to be the uncertainty of $\sigma(v)$ due to the lack of input data for temperatures $T>336$ K. To illustrate this, we added to the set of 10 FAMU experimental data
\cite{pla20,pla21} a fictitious 11-th data point $\Lambda_{11}=\Lambda(500)$ in the interval  
$8.\,10^{10}\le\Lambda_{11}\le13.\,10^{10}\text{ s}^{-1}$, obtained by extrapolating to 500 K the best fits shown on Fig.~1 of Ref.~\cite{ours}; the area spanned by the corresponding $\sigma(v)$-curves, obtained with the
``extended data sets'' is shadowed on Fig.~\ref{fig:symbols}. While by no means this should be considered as an evaluation of the systematical uncertainty of $\sigma(v)$, the plot
displays its fast exponential growth for $v\gtrsim5000$ m\,s$^{-1}$. We therefore set  $v=5000$ m\,s$^{-1}$ as upper limit, and   $v=200$ m\,s$^{-1}$ -- as lower limit of the range of validity of our computations of $\sigma(v)$. 

\subsection{Comparison with theory}
\label{sec:3B}

\begin{figure}[h]
\begin{center}
\includegraphics[scale=0.6]{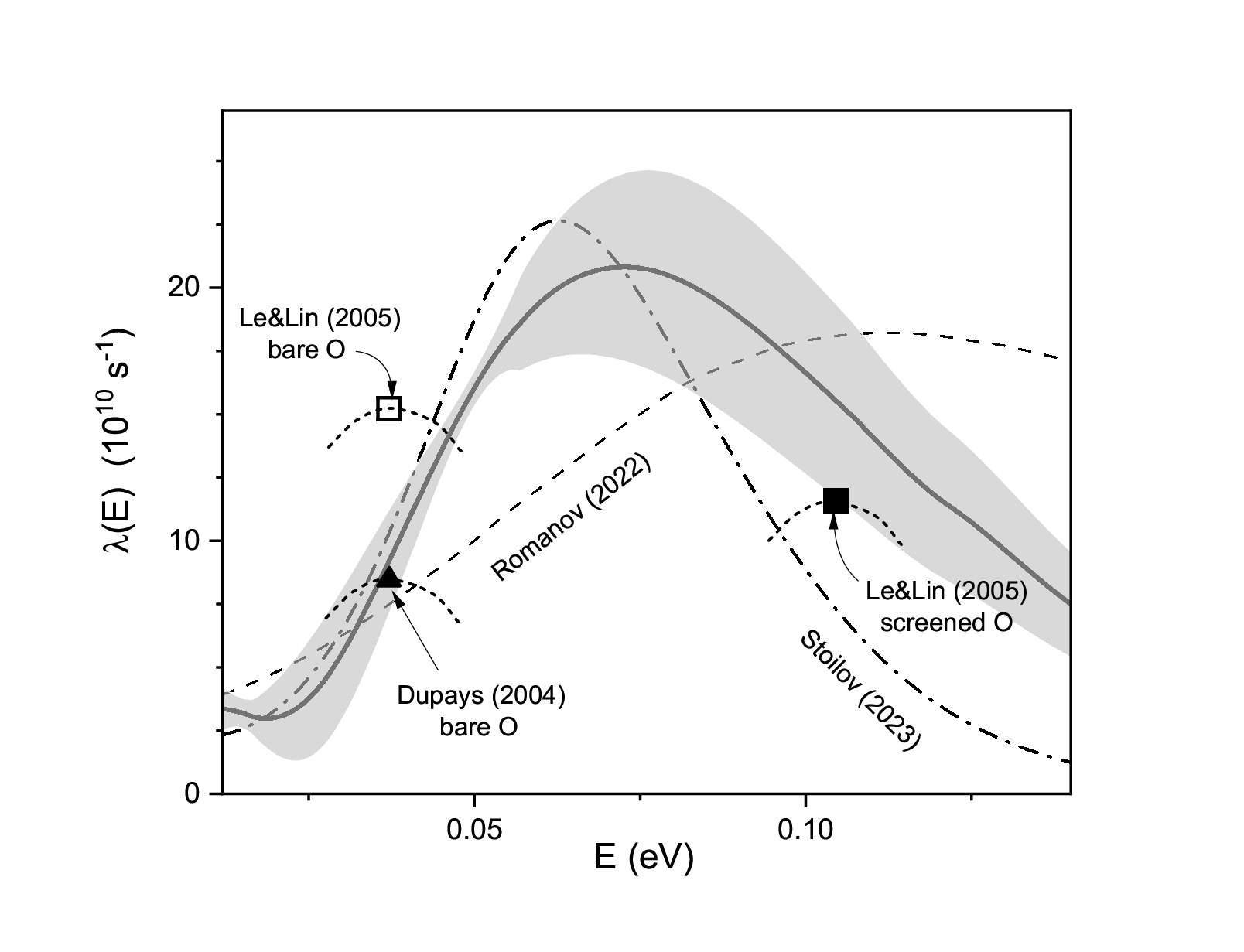}
\caption{Rate of muon transfer from muonic hydrogen to oxygen, obtained in the present work (solid), and its statistical uncertainty (shadowed area).  For comparison: the theoretical curve of Ref.~\cite{romanov22}, variant C (dashed), the local peak positions and values, calculated in \cite{dupays1,cdlin}, and the energy dependence determined from experiment in \cite{ours} without accounting for the oxygen molecular effects (dash-dotted). All rates are normalized to LHD.}
\label{fig:theories}
\end{center}
\end{figure}

Verification of the theoretical calculations of the muon transfer rate to oxygen was one of the motivations for the present paper. 
All theoretical works dedicated to the study of muon transfer to oxygen \cite{dupays1,cdlin,tcherbul,romanov22} consider reaction (\ref{eq:transfer}) as a three-body process involving the proton and muon of $p\mu$ and one of the nuclei of O$_2$; of the possible oxygen molecular effects only the electron screening of the Coulomb interparticle interaction potential was approximately accounted for in Ref.~\cite{cdlin}. 
To compare with theory, we evaluate the muon transfer rate $\lambda(E)$ as
\begin{align}
\lambda(E)=\rho_{\rm LHD}\,v\,\sigma(v)\vert_{v=\sqrt{2E/m_{A,N}}},
\end{align}
where $m_{A,N}=m_Am_N/(m_A+m_N)$ is the reduced mass of $p\mu$ and the oxygen nucleus.
Fig.~\ref{fig:theories} shows the plot of $\lambda(E)$ (the solid curve) and its statistical uncertainty corridor (the shadowed area). The dashed curve represents the energy dependence of the muon transfer rate reported in the most recent theoretical paper \cite{romanov22}, variant C. For the earlier papers we only mark the local peaks of the curves, calculated in \cite{dupays1} for a bare oxygen nucleus, and in \cite{cdlin} also with electron-screened Coulomb potential. Finally, the dashed-dotted curve represents 
$\lambda(E)$, obtained in \cite{ours} without accounting for the O$_2$-molecule degrees of freedom. 

\begin{figure}[h]
\begin{center}
\includegraphics[scale=0.6]{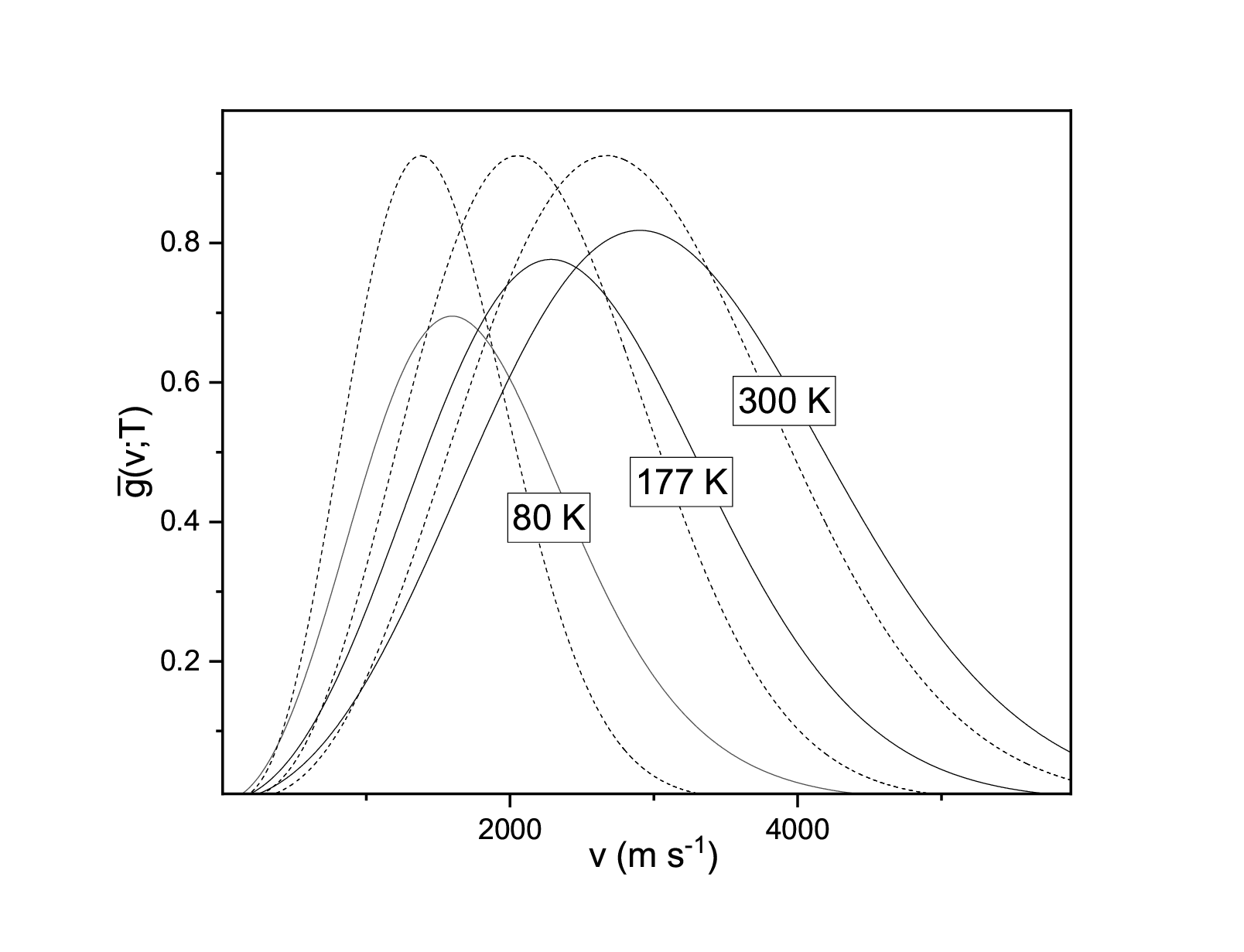}
\caption{Comparison of the kernels $\overline{g}(v;T)$, calculated for temperatures $T=80$, 
177, and 300 K with account of the oxygen nucleus motion described by the velocity distribution 
$\overline{f_N}$ (solid), and for frozen oxygen nuclei (dashed). }
\label{fig:efig}
\end{center}
\end{figure}

The results of the present work and of Ref.~\cite{ours} were obtained using the same set of experimental data as input and similar computational methods. The visible difference between them should be attributed to the effects of the oxygen nucleus motion that were not taken into account in \cite{ours}. This is clearly seen on Fig.~\ref{fig:efig}, where the kernels 
$\overline{g}(v;T)$ of Eq.~(\ref{eq:ofAM}) are juxtaposed to their counterpart for ``frozen'' oxygen nuclei (in which case there is no convolution with the nucleus velocity distribution 
$\overline{f_N}$.)
The comparison of the results on Fig.~\ref{fig:theories} shows that the molecular effects ``stretch'' the $\lambda(E)$-curve is  towards higher collision energies, and the peak position is shifted from 63 meV to 73 meV without any significant change of the peak value, thus significantly improving the agreement with the peak positions predicted in \cite{cdlin} (with screening) and \cite{romanov22}, and with variant C of \cite{romanov22} as a whole.

\subsection{Muonic hydrogen disappearance rate and the FAMU experiment}

\begin{figure}[h]
\begin{center}
\includegraphics[scale=0.6]{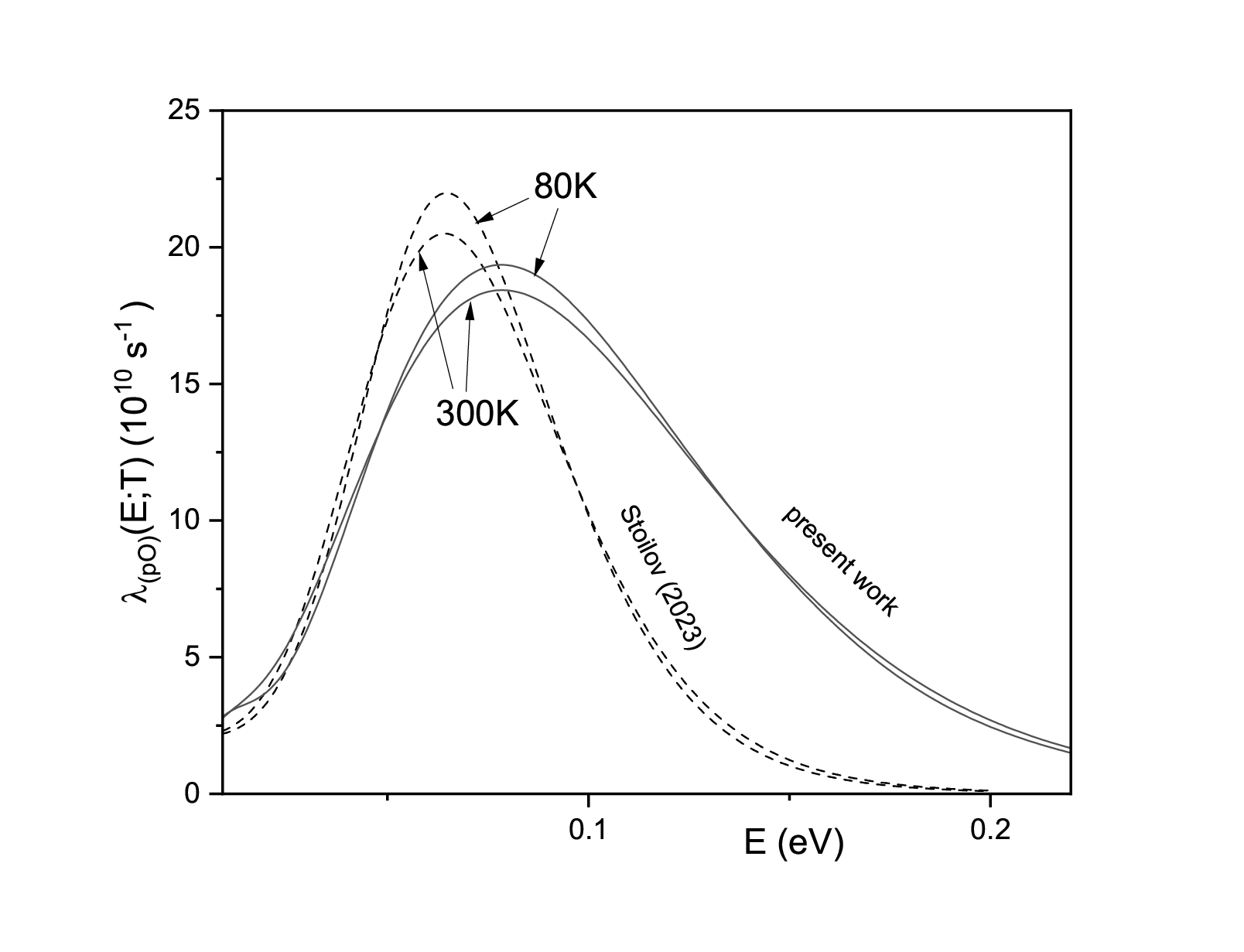}
\caption{Muon transfer contribution $\lambda^{(pO)}(E;T)$ to the disappearance rate 
$\lambda^{\rm(dis)}(E,T)$ of the $p\mu$ atoms for target temperature 80K and 300K (solid lines) and, for comparison, the rate of muon transfer calculated in \cite{ours} without account of the O$_2$ molecular structure (dashed). All rates are in units $10^{10}$ s$^{-1}$.}
\label{fig:disap}
\end{center}
\end{figure}

Knowledge of the energy dependence of the muon transfer to oxygen cross section allows to  calculate the muon transfer contribution $\lambda^{(pO)}(E;T)$ to the $p\mu$ disappearance rate, needed in Eq.~(\ref{eq:timespect}) for the evaluation of the characteristic X-ray time spectrum. 
For this we first calculate the rate of muon transfer to the nucleus of a thermalized oxygen molecule at temperature $T$ as function of the lab frame velocity $x$ of the $p\mu$ atom:
\begin{align}
 \lambda(x;T)&\!=\!4\pi\,x^2 \int d^3\mathbf{u}\,\int d^3\mathbf{v}\,v\,\sigma(u)
 \label{eq:disap80K}\\
&\times f_M(\mathbf{x}+\mathbf{v};T)\,f_N(\mathbf{v}-\mathbf{u};T)
\nonumber\\
&=\frac{1}{2x}\int_0^{\infty}dv\,v^2\, I(v;T)\int_{|x-v|}^{x+v}\frac{ds}{s}\overline{f}_M(s;T),
\nonumber\\
&I(v;T)\!=\!\frac{1}{2v}\!\int_0^{\infty}\!du\,u\,\sigma(u)
\int_{|v-u|}^{v+u}\frac{dt}{t}\overline{f}_N(t;T)
\nonumber
\end{align}
The expression for $\lambda^{(pO)}(E;T)$ as function of the $p\mu$ kinetic energy then reads:
\begin{align}
&\lambda^{(pO)}(E;T)=\lambda(x;T)\vert_{x=\sqrt{2E/m_A}}.
\end{align} 
Figure~\ref{fig:disap} shows the energy dependence of $\lambda^{(pO)}(E;T)$ for 
$T=80$K (the FAMU working temperature) and $T=300$K; the temperature effect is of the order of 7\%. For comparison we also plot the transfer rate, obtained in \cite{ours} without taking into account the O$_2$ molecular effects; evaluating the $p\mu$ disappearance rate with it instead of $\lambda^{(pO)}(E;T)$ may lead to wrong predictions for the characteristic X-ray time distribution. 

The numerical values of $\lambda^{(pO)}(E;T)$ are given in the Supplement \cite{suppl}.

\section{Conclusions}

The present paper is extending and upgrading the results of a previous work \cite{ours}, dedicated to the 
determination of the rate of low-energy muon transfer to oxygen from the experimental data of the FAMU collaboration. By consistently taking into account the internal degrees of freedom of the oxygen molecule, we now obtain more accurate results for the muon transfer cross section that improve the agreement with the theoretical calculations of the process, and develop an efficient computational technique for modeling the physical processes underlying the FAMU experiment for the measurement of the hyperfine splitting in muonic hydrogen and the determination of the Zemach radius of the proton.

\section{Appendix}
\label{sec:Appendix}

The goal of this Appendix is to obtain an estimate of the impact of the $p\mu$ atom on the population of the lower ro-vibrational states of the O$_2$ molecule without developing a consistent quantum mechanical description of the inelastic scattering of $p\mu$ by O$_2$ molecules that includes the rearrangement channel, which would be far beyond the scope of the present paper. 

Our approach is based on the assumption that the transitions between the O$_2$ levels provoked by the interaction with the $p\mu$ atoms precede the muon transfer. Indeed, considerations in the hyperspheric approach show that the muon transfer occurs at distances below $R_c=10^{-11}$ m \cite{cdlin,dupays1}, which is significantly shorter than the range of the Coulomb interaction of $p\mu$ with the oxygen nucleus. This interaction, in view of the electric neutrality of $p\mu$, is approximated at large distances $R$ with the potential $U_{\rm as}(R)$ of the time-averaged induced electric dipole moment of $p\mu$ in the Coulomb field of the oxygen nucleus with charge $Z$ (see \cite{LL}, \S77, p.289).
\begin{align*}
U_{\rm as}(R)=-\frac{9}{2}\,e^2 a_{\mu}^3 \,\frac{Z^2}{R^4},\ R\gg R_c,
\end{align*}
where $a_{\mu}=4\pi\epsilon_0\hbar^2/m_{p,\mu}e^2=0.284748\,10^{-12}$ m is the Bohr radius for muonic hydrogen, expressed in terms of the reduced mass 
$m_{p,\mu}=(m_{\mu}^{-1}+m_p^{-1})^{-1}$, and $Z=8$. 
Accordingly, we use an efficient potential $U_{\rm eff}(R)$, which asymptotically approaches $U_{\rm as}(R)$ at large $R$, but us regularized at short distances to suppress the fast growth of $U_{\rm as}$ for $R<R_c$, where the muon transfer channel is assumed to dominate, and also involves the effective charge $Z_{\rm eff}=6.2$ instead of $Z=8$, as long as the nuclear charge screening effects have been shown in \cite{cdlin} to be of  importance:
\begin{align}
U_{\rm reg}(R)=-U_0\,\frac{1}{(R^2+R_c^2)^2},\ \ 
U_0=\frac{9}{2}\,e^2 a_{\mu}^3 Z_{\rm eff}^2.
\label{eq:ureg}
\end{align}

We consider the process of inelastic scattering of a monochromatic beam of $p\mu$ atoms in the ground state with initial momentum $\mathbf{p}$ by an O$_2$ molecule in the ground vibrational state, initial rotational state $|lm\rangle$ and final rotational state $\langle J'm'|$ 
\begin{align}
(p\mu)_{\rm 1s}\mathbf{p}+O_2(Jm)\rightarrow(p\mu)_{\rm 1s}\mathbf{p}'+O_2(J'm')
\end{align}
and evaluate the cross section of the process, $\sigma_{Jm,J'm'}(p)$, in the Born approximation:
\begin{align}
\sigma_{Jm;J'm'}(p)&\!=\!\frac{2\pi}{\hbar}\!
\int \frac{d^3p'}{(2\pi)^3}\left|
\,|\langle J'm'|\langle\mathbf{p}'|U_{\rm reg}|Jm\rangle|\mathbf{p}\rangle\right|^2
\nonumber\\
&\times\delta((p'^2\!-p^2)/2m_A+E_{J'}\!-\!E_J),
\label{eq:melsqd}
\end{align}
where $\mathbf{p}'$ and $m_A=m_p+m_{\mu}$ are the momentum of the outgoing $p\mu$ and its mass, while $E_{J}$ and $E_{J'}$ are the rotational energy levels of the ground 
vibrational state of O$_2$. (Excitation of higher vibrational levels by thermalized $p\mu$ atoms can be neglected.)
It can be shown that the criteria for the applicability of the Born approximation (see \cite{LL}, \S126) are nearly fulfilled, at least to the degree to allow for the quantitative estimate of the considered phenomena. Using the explicit expressions for the potential $U_{\rm reg}$ and the approximate wave functions $\tilde{R}^{(0)}_{0,J}$ of 
O$_2$ derived in \cite{pettitt}, the matrix element in (\ref{eq:melsqd}) can be put in the form
\begin{align}
&\langle J'm',\mathbf{p}'|U^{\rm reg}(R)|Jm,\mathbf{p}\rangle\label{eq:matel1}\\
&=\sqrt{\pi}(2\pi)^2\,\sqrt{\frac{m_A}{p}}\,U_0\sum_{LM}(-i)^L
\sqrt{\frac{(2L+1)(2J+1)}{(2J'+1)}}
\nonumber\\
&\times C_{J0,L0}^{J'0}\,C_{Jm,LM}^{J'm'}\,
I_L(k)
\frac{e^{-R_ck}}{2R_c}\,Y_{LM}^*(\hat{\mathbf{k}}).\nonumber,
\end{align}
where $C_{Jm,LM}^{J'm'}$ are Clebsch-Gordan coefficients, 
$\mathbf{k}=(\mathbf{p}'-\mathbf{p})/\hbar$, $\mathbf{\hat{k}}=\mathbf{k}/|\mathbf{k}|$, 
$Y_{LM}$ are spherical harmonics, and $I_L(k)$ denotes the integral involving the spherical Bessel functions $j_L$:
\begin{align*}
I_L(k)=\int_0^{\infty} dx\,(\tilde{R}^{(0)}_{0,0}(x))^2\,j_L(kx/2).
\end{align*}
Inserting (\ref{eq:matel1}) into (\ref{eq:melsqd}), averaging with respect to $m$,
summing over $m'$ and integrating over the final $p\mu$ momentum $\mathbf{p}'$ leads to the following expression for the cross section $\sigma_{JJ'}(v)$ as function of the velocity 
$v=p/m_A$ of the incoming $p\mu$ atom:
\begin{align}
&\sigma_{JJ'}(v)=\frac{(2\pi)^3}{16\hbar^2}\,
\frac{U_0^2}{R_c^2}\,\frac{1}{v^2}
\sum\limits_{L=|J'-J|}^{J+J'}(2L+1)
\label{eq:sigma-fin}
\\
&\times\left(C_{J0,L0}^{J'0}\right)^2
\int\limits_{k_1(v)}^{k_2(v)}dk 
\,k\,e^{-2R_ck}
\,\left(I_L(k)\right)^2,
\nonumber\\
&k_{1,2}(v)=\frac{m_A}{\hbar} \left|v\mp\sqrt{v^2-2(E_{J'}-E_J)/m_A}\right|.
\nonumber
\end{align}

The values of $\sigma_{JJ'}(v)$ were calculated for  $0\le J,J'\le40$ on a 80-point grid of values $100\le v\le 8000$ m\,s$^{-1}$. Maximal are the cross sections for transitions with 
$\Delta J=|J'-J|=1$; for slow $p\mu$ atoms with $v<400$ m\,s$^{-1}$ the values of 
$\sigma_{J,J\mp1}(v)\sim10^{-16}$ cm$^2$ exceed by more than an order of the muon transfer cross sections, though for higher velocities $v>3000$ m\,s$^{-1}$ decrease to $10^{-18}$ cm$^2$. The decrease with $\Delta J$ is faster, especially for slow atoms. 

\begin{figure}[b]
\begin{center}
\includegraphics[scale=0.6]{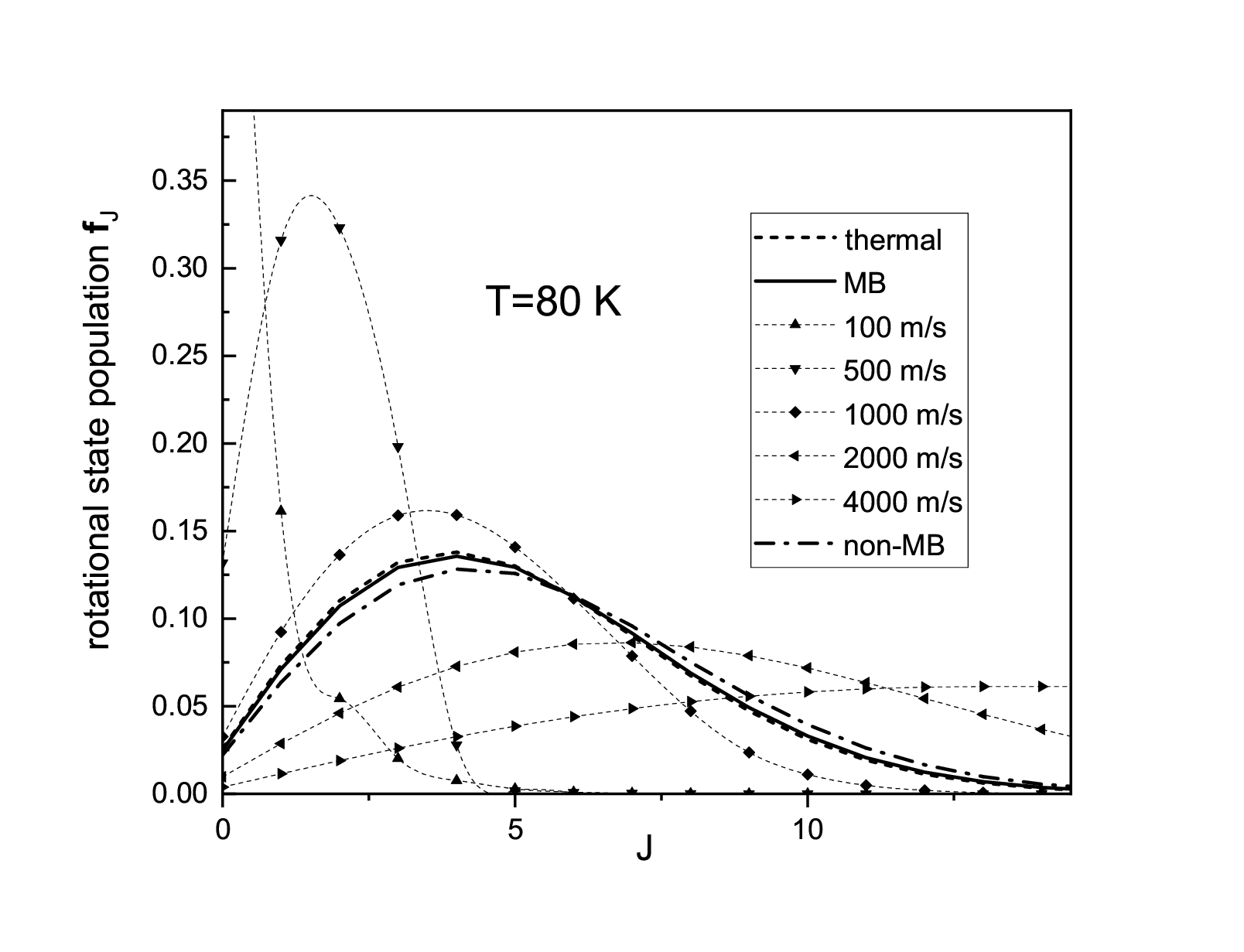}
\caption{Population $f_J$ of the rotational excitations $|J\rangle$ of the ground vibrational state of a thermalized oxygen molecule \underline{after} inelastic collision with a beam of muonic hydrogen atoms in ground state at 80 K. The thin dashed curves correspond to monochromatic beams of atoms propagating with the indicated velocity; the solid curve is for thermalized atoms with Maxwell-Boltzmann velocity distribution, and the dashed-dotted curve is for the non-Maxwell-Boltzmann distribution $\overline{f_A}$ of Eq.(\ref{eq:nst}). The thick dashed curve, hardly distinguishable from the solid one, is for the thermal population $f^{\rm th}_J$.}
\label{fig:avrat80}
\end{center}
\end{figure}

The large values of the collision-induced rotational transition cross sections require a careful evaluation of the resulting population $f_J, \sum_J f_J=1$ of the rotational states of O$_2$, since any deviation from the thermal population 
$f^{\rm th}_J$ will affect -- through the oxygen nucleus velocity distribution  
$\overline{f_{N}}$ and the kernel $\overline{g}(v;T)$ -- the extracted muon transfer cross section. The FAMU experiment is studying the muon transfer events from {\em thermalized} 
$p\mu$ atoms; the rate $\Lambda^{\rm rot}_{JJ'}$ of the rotational transition (normalized to LHD) is therefore averaged with the $p\mu$ velocity distribution 
\begin{align*}
\Lambda^{\rm rot}_{JJ'}=\rho_{\rm LHD}\int dv\,v\,\sigma_{JJ'}(v) \overline{f_A}(v)
\end{align*} 
and the needed final population $f_J$ will be given by
\begin{align*}
&f_{J'} = \sum_J (\exp\, \tau \mathbf{L})_{J'J} \,f^{\rm th}_J, \\
&\mathbf{L}_{J'J}=\Lambda^{\rm rot}_{JJ'}-\delta_{JJ'}\sum_{J''}\Lambda^{\rm rot}_{J'J''},
\end{align*}
where $\tau$ is a time interval of the order of the inverse muon transfer rate.
Fig.~\ref{fig:avrat80} shows that, as expected, scattering of fully thermalized muonic atoms has no impact on the already established thermal population of the oxygen molecule rotational levels; this was shown to hold for all temperature $70<T<336$ K. The deviation of the velocity distribution $\overline{f_A}$ from the Maxwell-Boltzmann distribution produces a visible effect, which is, however, suppressed in the convolution with $\overline{f_{AM}}$ in Eq.~(\ref{eq:ofAM}) and affects 
$\overline{g}(v;T)$ by less than 1\%. 

\end{document}